\documentclass[aps,pre,amsmath,amssymb,superscriptaddress,floatfix,showpacs,twocolumn]{revtex4}

\usepackage{graphicx}
\usepackage{bm}
\input epsf.sty

\begin{document}


\title{Characterizing steady state and transient properties of reaction-diffusion systems}

\author{Sven Dorosz and Michel Pleimling}
\affiliation{Department of Physics, Virginia Polytechnic Institute and State University, Blacksburg, Virginia 24061-0435, USA}

\date{\today}

\begin{abstract}
In the past the study of reaction-diffusion systems has greatly contributed to our understanding of the behavior
of many-body systems far from equilibrium. In this paper we aim at characterizing the properties of
diffusion limited reactions both in their steady states and out of stationarity.
Many reaction-diffusion systems have the peculiarity that microscopic
reversibility is broken such that their transient behavior can not be investigated through the study of most
of the observables discussed in the literature. For this reason we analyze the transient properties of
reaction-diffusion systems through a specific work observable
that remains well defined even in the absence of microscopic reversibility and that obeys an exact detailed fluctuation relation
in cases where detailed balance is fulfilled. 
We thereby drive the systems out of their nonequilibrium steady
states through time-dependent reaction rates.
Using a numerical exact method
and computer simulations,
we analyze fluctuation ratios of the probability distributions obtained during the forward and reversed processes. 
We show that 
the underlying microscopic dynamics gives rise to peculiarities in the configuration space trajectories,
thereby yielding prominent features in the fluctuation ratios.
\end{abstract}


\maketitle
\pagestyle{plain}
\section{Introduction}
Understanding general properties of systems that are far from equilibrium remains one of the most challenging 
problems in contemporary physics. In recent years some remarkable progress has been made in the study of
fluctuations in nonequilibrium small systems, see, e.g., \cite{Eva93,Eva94,Gal95,Jar97a,Jar97b,Kur98,Cro98,Leb99,
Cro99,Cro00,Jar00,Hat01,Muk03,Sei05,Spe05,Har07,Lip02,Col05,Dou05a,Wan02,Car04,Tie06,Jou08,Ber08}. This progress is mainly due to
the formulation of various exact fluctuation and work theorems that
provide very generic statements that hold true for large classes of systems. Some of the fluctuation theorems
deal with the rate of entropy production, either in systems that are in a nonequilibrium steady state \cite{Eva94,Gal95}
or in systems that initially are in equilibrium before being driven out of equilibrium by an external force \cite{Kur98,Leb99}.
One also distinguishes between detailed \cite{Cro99,Jar00,Sei05} and integral \cite{Jar97a,Hat01,Spe05} fluctuation theorems in cases
where the system is driven from one steady state to another in finite time.
In addition, work theorems \cite{Jar97a,Jar97b,Cro99} relate the free energy difference between two equilibrium states to the
amount of work done during the switching from one state to the other. Many of these theorems have been verified in various
experimental settings \cite{Lip02,Col05,Dou05a,Wan02,Car04,Tie06,Jou08,Ber08}, thus illustrating their usefulness for
characterizing nonequilibrium systems. 

In the past diffusion-limited reaction systems have been proven to be extremely useful in order to understand
the generic behavior of many-body systems far from equilibrium. Especially, the study of these systems is at the
origin of our understanding of the properties of nonequilibrium phase transitions \cite{Hen08,Odo08}.
Whereas phase transitions in reaction-diffusion systems are by now very well characterized, this is quite different
away from these special points. In the present work we discuss
different ways of characterizing the steady state and transient properties of generic reaction-diffusion systems,
thus contributing to a more comprehensive understanding of these important systems.

Recently discussed extensions
of exact fluctuation theorems to nonequilibrium systems with chemical reactions  
mostly focused on reversible reactions and reaction networks
\cite{Gas04,Sei04,And04,Sei05a,And06,Sch07,And08}.
However, many reaction-diffusion models are characterized by {\it irreversible} reactions,
thus yielding a breaking of the usually assumed microscopic reversibility. By breaking microscopic reversibility
we mean that if $\omega(C_i \longrightarrow C_j)$ is the transition probability from configuration $C_i$ to configuration $C_j$,
it can happen that $\omega(C_i \longrightarrow C_j) = 0$ even though $\omega(C_j \longrightarrow C_i) > 0$.
A direct consequence of the absence of microscopic reversibility is that many observables discussed
in the context of fluctuation theorems are then ill defined, as in their derivation one explicitly uses that 
for any path in configuration space the reversed path also exists \cite{Cho09,Ohk09}. For this reason we focus in our study of fluctuations
in reaction-diffusion systems on an observable that is well defined even in the absence of microscopic
reversibility. For a system initially in an equilibrium steady state this quantity
is identical to the work observable used in the Jarzynski and Crooks relations \cite{Jar97a,Jar97b,Cro00}.

It should be noted that diffusion-limited systems with effective irreversible reactions can be prepared
through a fast evacuation of some of the reaction products. This makes plausible a possible future verification
of the intriguing features that are revealed in our study.

In the following we discuss, using a numerical exact method and numerical simulations, steady state and
transient properties of various reaction-diffusion systems. Especially, we study fluctuations in 
systems that are initially in a steady state before being driven away from stationarity
by varying one of the reaction rates. This protocol allows us to measure
the probability distribution of our observable when going from a steady state $A$, characterized by the value
$r_A$ of some reaction rate $r$, to another steady state $B$, characterized by the value $r_B$ of the same
reaction rate. Defining the reversed process as changing the reaction rate backwards from $r_B$ to $r_A$, we can
measure also the probability distribution in that case and compare the distributions for the forward and reversed processes.
Even though no exact detailed fluctuation theorem is observed for the studied quantity in the absence of detailed balance, we show
that the fluctuation ratios display intriguing signatures due to the specific dynamics of the
nonequilibrium systems under investigation.

A brief account of some of our results has been given previously \cite{Dor09}. In the present paper we not only 
give a detailed study of the features observed in the ratio of the probability distributions
for our observable, we also extend our investigation to other reaction schemes not studied previously. 
This allows us to gain a better understanding of fluctuations in truly nonequilibrium systems driven away from stationarity
and to make the discussion in \cite{Dor09} more quantitative.

Our paper is organized in the following way. In the next Section we introduce the different reaction-diffusion models
and discuss their steady state properties. In Section III we drive these systems out of stationarity
through time-dependent reaction rates. Using an observable that remains well defined even in the absence of microscopic
reversibility, we study the probability distributions for this variable and show
that the fluctuation ratios formed by the probability distributions computed in the forward and the reversed processes
display features which can be related to the dynamical properties of the nonequilibrium system.
Finally, Section IV gives our summary
as well as an outlook on open problems.

\section{The models and their steady state properties}

We consider one-dimensional lattices made up of $L$ sites with periodic boundary conditions where every lattice site
can at most be occupied by one particle. Because of the exclusion of multiple occupancy of the lattice sites,
a total of $2^L$ configurations exist.
Particles are allowed to jump to unoccupied nearest 
neighbor sites with a diffusion rate $D$ and undergo various creation and annihilation reactions. 
We discuss in the following four basic reaction schemes, see Table \ref{table1}, and we denote with model 1, 2, 3, and 4 
the four models that result from these reaction schemes. In all four models we have an annihilation
process, that takes place with rate $\lambda$, as well as a creation process where a new particle is created with
rate $h$. The different models differ by the way creation and annihilation take place. Let us first discuss the
annihilation process. In models 1 and 2, two particles on neighboring sites undergo a reaction which leads to the destruction of one of the particles. This is different in model 3, where both particles are destroyed at the same time. Finally,
in model 4 three neighboring particles are destroyed in the annihilation reaction. For the
creation process we note that whereas in models 2, 3, and 4 new particles are spontaneously created at empty sites,
in model 1 a new particle can only be created at an empty site if one of the neighboring sites is already occupied.

\begin{table}[thb]
\begin{tabular}{|c|c|c|c|}
\hline
model 1 & model 2 & model 3 & model 4\\
\hline
$A+A\stackrel{\lambda}{\to} 0+A$ & $A+A\stackrel{\lambda}{\to} 0+A$ & $2A\stackrel{\lambda}{\to} 2 \, 0$
& $3A\stackrel{\lambda}{\to} 3 \, 0$\\
$A+0 \stackrel{h}{\to} A+A $& $0\stackrel{h}{\to} A $ & $0\stackrel{h}{\to} A $ & $0\stackrel{h}{\to} A $\\
\hline
& model 2' & model 3' & model 4'\\
\hline
& $A+A\overset{\lambda}{\underset{\varepsilon_\lambda\lambda}{\rightleftarrows}} 0+A$ &
$2A \overset{\lambda}{\underset{\varepsilon_\lambda\lambda}{\rightleftarrows}}  2 \, 0 $  &
$3A \overset{\lambda}{\underset{\varepsilon_\lambda\lambda}{\rightleftarrows}} 3 \, 0$\\
& $0 \overset{h}{\underset{\varepsilon_hh}{\rightleftarrows}} A $ &
$0 \overset{h}{\underset{\varepsilon_hh}{\rightleftarrows}} A$ &
$0 \overset{h}{\underset{\varepsilon_hh}{\rightleftarrows}} A$ \\
\hline
\end{tabular}
\caption{The different reaction schemes discussed in this work. Whereas model 1 is an equilibrium model,
in models 2, 3, and 4 microscopic reversibility is partly or fully broken.
In the modified models 2', 3' and 4' we allow for reversible reactions with rates $\varepsilon_hh$ and $\varepsilon_\lambda\lambda$,
with $0 < \varepsilon_h \leq 1$ and $0 < \varepsilon_\lambda \leq 1$.} \label{table1}
\end{table}

These creation and annihilation processes have been chosen in such a way that the models present different degrees of microscopic 
reversibility. Thus in model 1 all reactions are reversible, and it is easy to see that this model
is in chemical equilibrium for fixed values of the reaction and diffusion rates. Indeed, if both $\lambda$ and $h$ are
different from zero, the reaction scheme reduces to a unique reversible reaction and detailed balance is fulfilled.
Model 2, on the other hand, does allow for some reactions to be irreversible. For example, 
a new particle can be created in the middle of two empty sites, $000 \longrightarrow 0A0$, with rate $h$, but it is
not possible to go back immediately to three empty sites as the newly created particle needs a neighbor for the annihilation
process to take place. Finally, all reactions are irreversible in models 3 and 4, yielding a complete absence of microscopic reversibility.

We also studied variants of model 2, 3, and 4, called 2', 3', and 4', where we restore microscopic reversibility by allowing the reversed processes to take place with rates $\varepsilon_hh$ and $\varepsilon_\lambda\lambda$, where
$0 < \varepsilon_h$, $\varepsilon_\lambda \leq 1$. Even though all reactions are now reversible, these models do not
fulfill detailed balance for $\varepsilon_h$, $\varepsilon_\lambda < 1$ and are therefore still nonequilibrium models.

For all our models the dynamics is described by a discrete-time master equation for the probability $P(C_i,t)$ that the system is in
configuration $C_i$ at time $t$ \cite{Gaspard04}:
\begin{eqnarray}
P(C_i,t+1) - P(C_i,t) & = & \sum\limits_{j} \left[ \omega(C_j \longrightarrow C_i) \, P(C_j,t) - \right. \nonumber \\
&& \left. \omega(C_i \longrightarrow C_j) P(C_i,t) \right] ~.
\end{eqnarray}
Zia and Schmittmann \cite{Zia06,Zia07} pointed out that for this type of systems 
a nonequilibrium steady state is characterized by both the stationary
probability distribution $P_s(C_i)$ and the stationary probability currents
\begin{equation}
K^*(C_i,C_j) = \omega(C_j \longrightarrow C_i) \, P_s(C_j) - \omega(C_i \longrightarrow C_j) P_s(C_i)
\end{equation}
between two configurations $C_i$ and $C_j$. 

The stationary probabilities are readily obtained for fixed reaction and diffusion rates by setting up
the transition probability matrix {\bf W} whose elements are the 
transition rates between different configurations. Of course, as we have $2^L$ configurations
for a system of size $L$, the transition probability matrix is a $2^L \times 2^L$ matrix. The stationary probabilities
are then obtained as the elements of the null eigenvector of the Liouville matrix {\bf L} which results when
subtracting off the identity matrix from the transition probability matrix. For systems that are small enough this eigenvalue problem can be solved using standard algorithms. For larger system sizes, the stationary probabilities can be measured through standard Monte Carlo simulations.

\begin{figure}[ht]
\centerline{\epsfxsize=3.00in\ \epsfbox{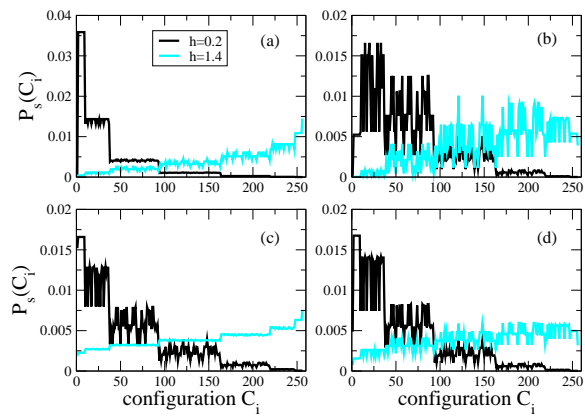}}
\caption{(Color online) The stationary probabilities for (a) model 1, (b) model 2, and (c,d) model 3. 
The common parameter is $\lambda=1$, whereas $D=1$ in (a,b,c) and $D=5$ in (d). The system size is $L=8$.
The configurations are grouped by the total number of particles in the system, with the empty configuration on the 
left and the fully occupied lattice on the right.}
\label{fig1}
\end{figure}

We show in Fig. \ref{fig1} the stationary probability distributions for some of the models and
various values of the reaction rates. Configurations with the same number of particles are grouped together, 
with the empty configuration to the left and the fully occupied lattice to the right. 
The first thing to note is that a change of reaction rates has a large impact on the stationary probability
distributions. When the creation of new 
particles takes place with a small rate, see the black lines in Fig. \ref{fig1},
configurations with only few particles are the most likely.
This is different when the creation rate is large, as then
configurations with a large number of occupied sites have an increasing weight,
see the cyan (light gray) lines in Fig. \ref{fig1}. A corresponding behavior is observed when 
changing the rate $\lambda$. On the other hand, however,
a change of the value of the diffusion constant $D$ mainly changes the distributions 
quantitatively, see Fig. \ref{fig1}(c) and (d). 

Even though there are visible differences in the 
stationary probability distributions between Fig. \ref{fig1}(a) and (c), it is not possible to guess from these
stationary probabilities alone whether the system is in equilibrium, as it is the case for model 1, or whether we are dealing
with a nonequilibrium system with a fully irreversible reaction scheme, as for model 3.
The stationary probability distribution does not allow by itself to characterize unequivocally nonequilibrium steady states.

\begin{figure}[ht]
\epsfxsize=3.00in\ \epsfbox{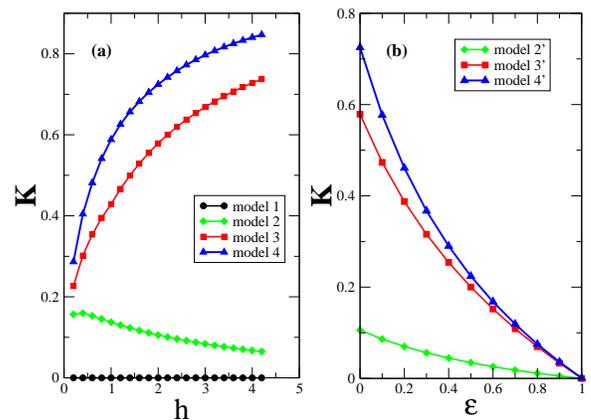}
\caption{(Color online) The total probability current $K$ (a) as a function of the creation rate $h$ for models 1, 2, 3, and 4,
and (b) as a function of $\varepsilon=\varepsilon_\lambda=\varepsilon_h$ for models 
2', 3', and 4'. In all cases $\lambda=1$ and $D=1$.
In (b) the creation rate is $h=2$. The data are for systems with $L=8$ lattice sites.}
\label{fig2}
\end{figure}

Instead of analyzing one-by-one the stationary probability currents for various configuration pairs, it is more convenient
to look at the global quantity \cite{Pla09}
\begin{equation}\label{eq:K}
K = \sum\limits_{i,j;\, i< j}
|K^*(C_i,C_j)|~.
\end{equation}
In Fig. \ref{fig2}(a) we show the dependence of $K$ on the value of the creation rate $h$ for fixed values
of $\lambda$ and $D$. Obviously, the dependence is very different for the different models. In the equilibrium
model 1 one does not have any non-vanishing stationary probability currents, and $K$ is zero for all values of the
reaction and diffusion rates, as expected. This is different for the nonequilibrium systems which are characterized by
non-vanishing stationary probability currents. Interestingly, the value of $K$ decreases in model 2 for larger $h$, whereas for models 3 and 4 it increases as a function of $h$. In order to understand this difference in behavior, we recall that for larger values of $h$ configurations with a large number of particles have an increased stationary probability.
As a result, free sites will have with high probability occupied neighboring sites, 
and the creation process $0\to A$ effectively equals the process $A\to 2A$. This is, however, exactly the reversed
reaction to the annihilation process of model 2, which explains why for large $h$ the behavior of model 2 approaches
that of an equilibrium system. For models 3 and 4, however, all reactions remain irreversible and $K$ keeps on growing.

In Fig. \ref{fig2}(b) we plot $K$ as a function of the parameter $\varepsilon=\varepsilon_\lambda=\varepsilon_h$ for the
modified models 2', 3', and 4' for which we allow the reversed reactions to take place. As a result, all reactions
are now reversible, but for all cases we are still out of equilibrium as long as $\varepsilon < 1$. Increasing
$\varepsilon$ decreases the distance to equilibrium which is finally reached for $\varepsilon =1$ when all reactions and the corresponding reversed reactions are taking place with the same rate.

This discussion of the stationary probability distributions and of the stationary probability currents
shows that it is not possible to characterize in an unambiguous way a nonequilibrium system solely 
through its stationary probability distribution.
Much information is contained in the stationary probability currents which do allow to distinguish between the properties
of equilibrium, weakly nonequilibrium, and strongly nonequilibrium systems.
Therefore these currents allow to quantify the distance to equilibrium, making them a useful tool for the
characterization of nonequilibrium systems. 

\section{Transient fluctuation relations}
Having discussed the steady-state properties of reaction-diffusion models, we now focus on the
characterization of the transient behavior when the systems are brought out of stationarity and are then
allowed to relax to a new steady state. We are realizing this through a protocol in which we
change one of the reaction rates. Experimentally, a change of rates of chemical reactions can
be achieved by changing the temperature, for example. In our protocol we change one of the rates $r$
from an initial value $r_0$ to a final value $r_M$ in $M$ equidistant steps of length $\Delta r$,
yielding for the reaction rate the values
$r_i = r_0 + i \Delta r$ with $i=0,\cdots , M$. We assume that
at every step only one reaction or diffusion process takes place.

In the following we discuss mainly numerical exact results for small one-dimensional systems. 
This numerical exact approach is rather straightforward and is summarized in the Appendix.
Larger systems can be studied along the same lines through numerical simulations, but 
this must be done with some care in order to guarantee a sufficient sampling of rare events \cite{Dor10}.

\subsection{Observables}

We discuss in the following two observables which differ by the fact that for one of the variables
microscopic reversibility has to be assumed whereas for the other no such assumption has
to be made. For a system that is microscopic reversible,
as for example a system that fulfills detailed balance, we will discuss the difference between these two
quantities explicitly.

In order to define these quantities let us first suppose that the system is in a steady state.
Starting from a configuration $C_0$, the system is in the configuration $C_i$ at step $i$,
such that after $M$ steps the system has performed the following path in configuration state:
${\bf X} = C_0 \longrightarrow C_1 \longrightarrow \cdots
\longrightarrow C_{M-1} \longrightarrow C_M$. The probability for this path is 
\begin{equation}
P\left( {\bf X} \right) = P_s(C_0) \prod\limits_{i=0}^{M-1} \omega(C_i \longrightarrow C_{i+1} )~,
\end{equation}
where $\omega(C_i\to C_{i+1})$ is the transition probability from configuration $C_i$ to configuration $C_{i+1}$.
Denoting the reversed path by ${\bf \tilde{X}} =
C_M \longrightarrow C_{M-1} \longrightarrow \cdots \longrightarrow C_1 \longrightarrow C_0$, one then 
defines for Markovian systems the quantity \cite{Leb99,Sei05a}
\begin{equation}\label{ds_ss}
R_{ss}=\ln \frac{P({\bf X})}{P({\bf \tilde X})} =
\ln \frac{P_s(C_0)}{P_s(C_M)}+\sum_{i=0}^{M-1} \ln \frac{\omega(C_i\to C_{i+1})}{\omega(C_{i+1}\to C_{i})} ~.
\end{equation}
When the system is driven out of stationarity, we can generalize this definition to a time dependent
reaction rate, yielding
\begin{equation}\label{ds}
R =\ln \frac{P_s(C_0,r_0)}{P_s(C_M,r_M)}+\sum_{i=0}^{M-1} \ln 
\frac{\omega(C_i\to C_{i+1},r_{i+1})}{\omega(C_{i+1}\to C_{i},r_{i})} 
\end{equation}
where $P_s(C_i,r_i)$ is the probability to find the configuration $C_i$ in the stationary state
corresponding to the value $r_i$ of the reaction rate $r$ and $\omega(C_i\to C_{i+1},r_{i+1})$ is the 
transition probability from $C_i$ to $C_{i+1}$ at step $i+1$. 

A closer look at the observable $R$ reveals that its definition requires that
if $\omega(C_i\to C_{i+1},r_{i+1}) > 0$ than $\omega(C_{i+1}\to C_{i},r_{i})$ also has to be non zero. However,
in some of our reaction-diffusion models this condition is not fulfilled as microscopic reversibility is broken, 
and we can not use $R$ to study them.
Hatano and Sasa \cite{Hat01} have proposed a different quantity that is closely related to $R$, but
that does not assume microscopic reversibility. Adapting this quantity for systems driven out of stationarity,
we can write it in the following way \cite{Dor09}
\begin{equation}\label{dphi}
\phi = \sum\limits_{i=0}^{M-1} \ln \left[ \frac{P_s(C_i , r_i)}{P_s(C_i , r_{i+1})} \right]~.
\end{equation}
The quantity $\phi$ has been called the driving entropy production in \cite{Esp07}.

For a system with microscopic reversibility we can derive a relation between $R$ and $\phi$.  With the help
of the probability current
\begin{eqnarray}
K^*(C_i, C_{i+1}, r_{i+1})& =& \omega(C_{i+1} \to C_i, r_{i+1}) P_s(C_{i+1}, r_{i+1})- 
\nonumber \\
&&  \omega(C_i \to C_{i+1}, r_{i+1}) P_s(C_i, r_{i+1}) 
\end{eqnarray}
we can write Eq. (\ref{dphi}) for $\phi$ in the following form:
\begin{eqnarray} \label{Rphi}
\phi & = & R - \sum_{i=0}^{M-1} \ln \left[ - \frac{ K^*(C_i, C_{i+1}, r_{i+1})}{P_s(C_{i+1},r_{i+1}) \, 
\omega(C_{i+1} \to C_i, r_i)} \right. \nonumber \\
&& \left. +  \frac{ \omega(C_{i+1} \to C_i, r_{i+1})}{\omega(C_{i+1} \to C_i, r_i)} \right]
\end{eqnarray}
which reveals that the difference between $R$ and $\phi$ is composed of terms which have very different physical origins.
The first term in the $\ln$ in Eq. (\ref{Rphi}) is due to non-vanishing probability currents between different
configurations and is therefore characteristic for nonequilibrium states. The second term
is non-trivial only in transient processes as it accounts for a shift in the reversed transition probability.
This term reduces to the trivial value 1 in case one 
remains in a given steady state, with $r_{i+1} = r_i = r_0$ for all $i$.
If this steady state is in addition an equilibrium state, the probability currents are all vanishing, and one has $R=\phi=0$. 

It is easy to show \cite{Hat01,Sei05,Esp07} that for transient processes both quantities fulfill an integral fluctuation theorem: $\langle
e^{-R} \rangle = 1$ and $\langle e^{-\phi} \rangle =1$, 
where the average is taken over all possible histories when driving the system out of a general steady state. 
For a system that is initially in an equilibrium steady state the relation
$\langle e^{-\phi} \rangle =1$
reduces to the Jarzynski relation \cite{Jar97a} as then $\phi = \beta (W - \Delta F)$, where $W$ is the
work done on the system, $\Delta F$ is the free energy difference between initial and final states, and 
$\beta$ is the inverse temperature. The difference $W_d = W - \Delta F$ is the dissipative work.

\subsection{Probability distributions}

In systems with detailed balance, an exact fluctuation relation is obtained when plotting the ratio between the 
probability distributions 
$P_F(\beta W_d)$ and $P_R(-\beta  W_d)$ of the dissipative work $W_d$ done on the system in the forward and reversed processes \cite{Cro00}:
\begin{equation} \label{eq:crooks}
\frac{P_F(\beta W_d)}{P_R(-\beta W_d)} = e^{\beta W_d}~.
\end{equation}
Recalling that for systems with detailed balance we have the identity $\phi = W_d$, it is tempting to ask whether for $\phi$
an exact fluctuation theorem like (\ref{eq:crooks}) can also be encountered for a system initially in a nonequilibrium steady 
state. In fact, this is not the case: the absence of detailed balance in a nonequilibrium steady state 
entails non-zero probability currents, and no simple relation like the relation (\ref{eq:crooks}) exists for $\phi$ in this case.
As we shall discuss below, the corresponding fluctuation ratios yield {\it systematic} deviations from the simple behavior encountered
in systems with detailed balance, these deviations containing non-trivial information on the nonequilibrium system at hand.

However, before analyzing these ratios of probability distributions, we shall first discuss the probability distributions
themselves.

\begin{figure}[ht]
\epsfxsize=3.00in\ \epsfbox{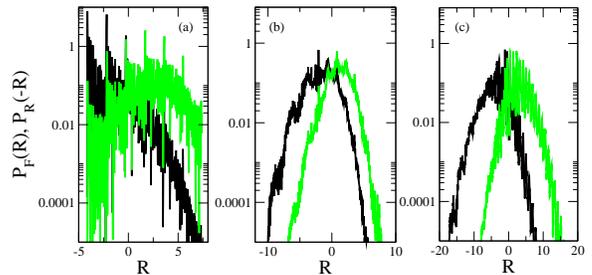} 
\caption{(Color online) Probability distributions for the quantity $R$ when the creation rate is changed in M=6 
equidistant steps from 0.2 to 1.4 ($P_F(R)$, black curve) or from 1.4 to 0.2 ($P_R(-R)$, green (gray) curve). The data have 
been obtained for a system
with $L = 8$ sites, with $D=5$ and $\lambda = 1$. (a) Model 1, (b) model 2' with $\varepsilon= 0.1$, and (c)
model 3' with $\varepsilon= 0.1$.}
\label{fig3}
\end{figure}

\begin{figure}[ht]
\epsfxsize=3.00in\ \epsfbox{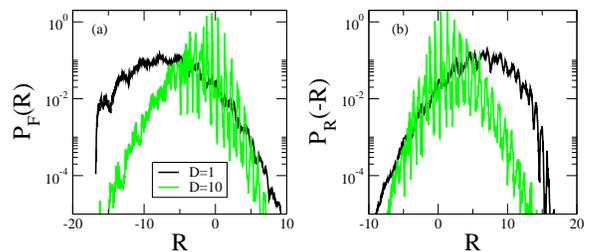}
\caption{(Color online) The same as in Fig. \ref{fig3}, but now for model 3' and two different values of the 
diffusion rate. (a) $P_F(R)$ from the forward process and (b) $P_R(-R)$ from the reversed process.}
\label{fig4}
\end{figure}

\begin{figure}[ht]
\epsfxsize=3.00in\ \epsfbox{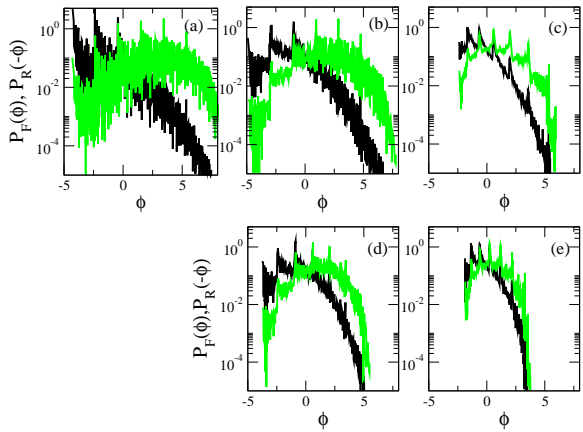}
\caption{(Color online) Probability distributions for the quantity $\phi$ when the creation rate $h$ is changed in M=6 steps 
from 0.2 to 1.4 ($P_F(\phi)$, black curve) or from 1.4 to 0.2 ($P_R(-\phi)$, green (gray) curve). The data have been obtained for a system 
with $L = 8$ sites, with $D=5$ and $\lambda = 1$. (a) Model 1, (b) model 2, (c) model 3, (d) model 2' with $\varepsilon= 0.1$, 
and (e) model 3' with $\varepsilon= 0.1$.}
\label{fig5}
\end{figure}

\begin{figure}[ht]
\epsfxsize=3.00in\ \epsfbox{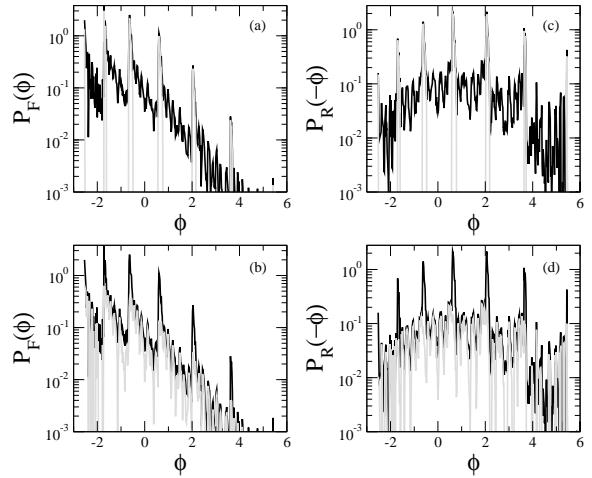}
\caption{(Color online) Main contributions to the probability distributions for $\phi$ in the forward and reversed processes.
The black lines show the full probability distributions whereas the gray lines show the contributions coming
from (a,c) trajectories in configuration space with only diffusion steps and no reactions and (b,d) from 
trajectories where exactly one reaction takes place that changes the number of
particles in the system.  
The data are for model 3 with $D=10$, $h_0=0.2$, $\Delta h=1.2$, and $\lambda =1$. 
The system size is $L=8$ and the driving length is $M=6$. }
\label{fig6}
\end{figure}

Figures \ref{fig3}-\ref{fig6} show typical examples for the probability distributions of $R$ and $\phi$ when changing the
creation rate from an initial value $h_0$ to a final value $h_M$ 
in $M$ steps (we only show the case of a varying creation rate $h$, but 
the following discussion can be made along similar lines when changing the value of the annihilation rate $\lambda$).

Fig.~\ref{fig3} shows the probability distributions of $R$ for three cases that fulfill  microscopic reversibility:
models 1, 2', and 3'. 
These different probability distributions are not Gaussian but are characterized by a rather irregular structure. 
Their shape depends on the dynamics of the different models, expressed by the different reaction schemes.
It is, however, not straightforward to relate specific features of the probability distributions to
the different reactions.
It is important to note that the peaks dominating these distributions do not have their origin
in the noisiness of some numerical data, but are real as
we are using a numerically exact method. In addition, our numerically  exact method also allows us to circumvent
any issues that might appear due to an insufficient sampling of rare events. This will be of importance in the
next section when we discuss the ratios of the forward and reversed probability distributions.\\

The probability distributions show a strong dependence on the system parameters.
This is illustrated in Fig.~\ref{fig4} where we compare for model 3' the distributions obtained for different values 
of the diffusion rate.
When we increase the diffusion rate, the general shape of the probability distribution changes and,
in addition, a large number of distinct peaks appear.

The probability distributions for $\phi$ differ markedly from those for $R$, see Fig. \ref{fig5}. This was expected
as the main difference between both quantities are the probability currents which are non-zero for a system
that is out of equilibrium. It is only for the equilibrium model 1 that the distributions for both quantities are
similar. Interestingly, the probability distributions for $\phi$ for both the forward and reversed
processes are characterized by the presence of prominent peaks. An increase of the diffusion constant strongly amplifies these
peaks but does not change the overall shape of the probability distributions. The fact that the heights of the peaks depend
on the value of the diffusion constant 
indicates that these peaks are related to trajectories in configuration space that are dominated by
diffusion steps and not by reactions. In Fig. \ref{fig6} we verify for model 3 that the main
contributions to the peaks for a drive of length $M =6$ indeed come from the trajectories where only diffusion takes place
such that the number of particles is constant along these trajectories. The subleading contribution, also
shown in Fig. \ref{fig6}, comes from the trajectories where a single reaction takes place which changes the number
of particles in the system. Because the peaks are dominated by trajectories with pure diffusion, the positions of
the peaks are the same for the forward and reversed processes, the leftmost peak resulting from the diffusion of a single
particle in the system, whereas the rightmost peak is due to the diffusion of a single empty site in the system.

Before closing this section, we remark that in \cite{Esp07}
similar peaks have been observed in the probability distributions of the driving entropy
production as well as of other related quantities in a model for electron transport
through a single level quantum dot.

\subsection{Fluctuation ratios}

\begin{figure}[ht]
\epsfxsize=3.00in\ \epsfbox{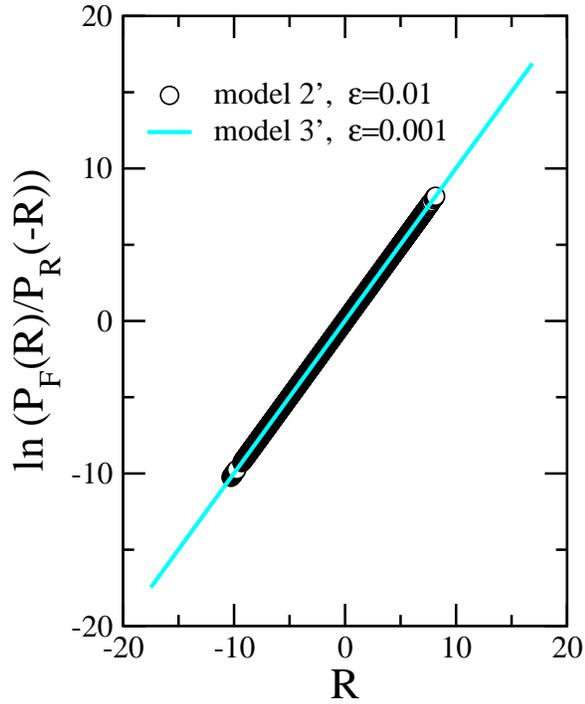}
\caption{(Color online) Fluctuation relation for the observable $R$ for model 2' and model 3' for different values of the parameter $\varepsilon$. 
The parameters in this calculation are $h_0=0.2$, $\Delta h=1.2$, $\lambda=1$, and $D=5$. The system size is $L=8$ and the 
driving length is $M=6$.}\label{fig7}
\end{figure}

Having just discussed the probability distributions of the quantities $R$ and $\phi$, we now
move on and study the fluctuation ratios formed by these probability distributions. 
For a system driven out of an initial equilibrium state and fulfilling detailed balance,
Crooks has shown the exact relation (\ref{eq:crooks}) to exist between the probability distributions
of the dissipative work measured in the forward and time-reversed processes. This remarkable result can
be extended to systems that are still reversible microscopically but that do not fulfill detailed balance any more 
\cite{Har07}. As illustrated in Fig. \ref{fig7} for models 2' and 3', the ratios of the probability distributions
for $R$ show a simple exponential dependence on $R$. The perfect exponential obtained from our data nicely validates
our numerical exact approach. Obtaining a plot of similar quality through Monte Carlo simulations is difficult as
rare events are then hard to measure.

\begin{figure}[ht]
\epsfxsize=3.00in\ \epsfbox{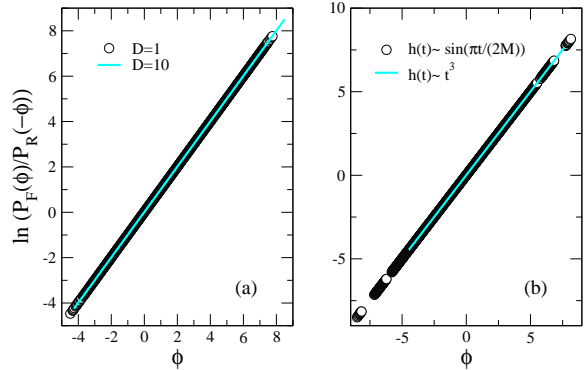}
\caption{(Color online) Fluctuation relation for the observable $\phi$ for model 1 with (a) different values of $D$ and (b) 
different ways of changing the parameter $h(t)$ with $D=1$. The driving process usually studied in this paper and
which yields the data shown in (a) is $h(t) \sim t$. The parameters used in these calculations are $h_0=0.2$, 
$\Delta h=1.2$, and $\lambda =1$. The system size is $L=8$ and the driving length is $M=6$.}
\label{fig8}
\end{figure}

Even though in the absence of microscopic reversibility $R$ is ill defined, this is different for
$\phi$ as this quantity exclusively involves the steady-state probabilities,
see Eq. (\ref{dphi}). For an equilibrium system $\phi$ fulfills an exact fluctuation theorem as it then
reduces exactly to the dissipative work. As shown in Fig. \ref{fig8} for model 1, an exponential relation is indeed
obtained for all parameter values as well as for different driving processes $h(t)$.

\begin{figure}[ht]
\epsfxsize=3.00in\ \epsfbox{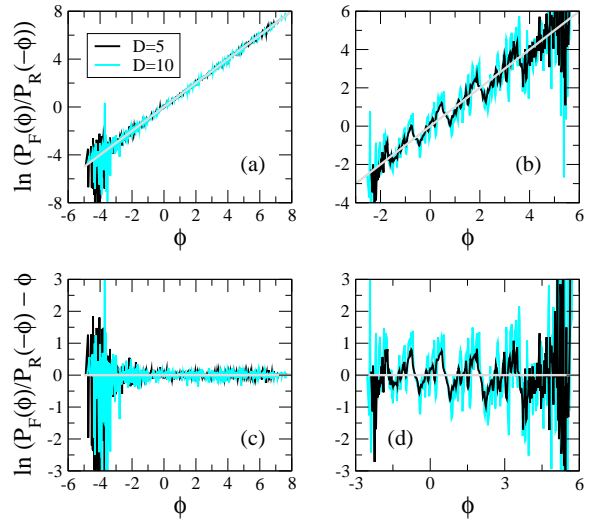}
\caption{(Color online) Fluctuation ratios for the observable $\phi$ for (a) model 2 and (b) model 3 and different values of 
the diffusion constant $D$. Whereas in model 2 only random deviations from a simple exponential behavior are
observed, systematic deviations show up for model 3. This is highlighted in (c) and (d) where we subtract $\phi$ 
from the logarithm of the fluctuation ratio. The light gray lines indicate a simple exponential dependence.
The parameters used in this calculation are $h_0=0.2$, $\Delta h=1.2$, 
and $\lambda =1$. The system size is $L=8$ and the driving length is $M=6$.}
\label{fig9}
\end{figure}

However, for a system with nonequilibrium steady states no exponential detailed fluctuation relation is expected for $\phi$ as this 
quantity does not contain the information on nonequilibrium currents, see Eq.\ (\ref{Rphi}). We show in Fig.\
\ref{fig9} ratios of the probability distributions of $\phi$ for models 2 and 3. For model 2 the deviations from the exponential 
are random and no pronounced dependence on system parameters, as for example the diffusion rate $D$, is observed.
For model 3, however, a qualitatively different behavior is encountered and {\it systematic} deviations
in the form of oscillations are observed. Similar oscillations are also observed for model 4 where three neighboring
particles are destroyed in the annihilation process. Interestingly, the amplitudes of these oscillations increase for increasing 
diffusion rates. At first one might think that this increase in peak height when increasing $D$
should be related to the increase of the peaks in the probability distributions themselves, see the discussion
in the previous section. However, this is too simplistic as an increase of peak heights in the 
probability distributions is also observed for models 1 and 2 for which we do not observe the corresponding
behavior in the fluctuation ratios. What is different between models 1 and 2 on the one hand and models 3  and 4 on the
other hand is that for the former models any change in the forward and reversed probability
distributions is compensated when forming the ratio (this compensation is exact for model 1 and approximate for
model 2), whereas
for the latter models this compensation is only partial, such giving rise to peaks also in the fluctuation ratios.

\begin{figure}[ht]
\epsfxsize=3.00in\ \epsfbox{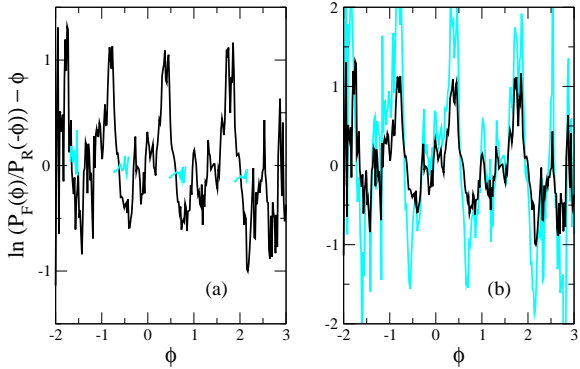}
\caption{(Color online) Comparison for model 3
of the fluctuation ratio (black line) with the ratio $\Pi_F(\phi)/\Pi_R(-\phi)$ (cyan (light gray) line) where $\Pi(\phi)$ is the probability
distribution of $\phi$ for all trajectories with (a) only diffusion steps and (b) exactly one reaction process.
Note that for trajectories with only diffusion few values of $\phi$ can be realized.
The common parameters are $h=0.2$, $\lambda=1.$, $M=6$ and $L=8$ and $D=5$.}
\label{fig10}
\end{figure}

Before discussing the origin of this difference, let us first have for model 3 a closer look at the peaks in the fluctuation ratio.
We first note that the positions of these peaks are {\it not} identical to
the positions of the extrema in the probability distributions (see for example Fig. \ref{fig6}). In Table \ref{table2}
we compare the positions of the maxima and minima in the fluctuation ratio with the peak positions in the probability
distributions. The observed offset means that the peaks in the probability distributions for the forward and reversed
processes compensate each other when forming the ratio, but that the compensation is only partial away from the peaks.
Recalling that the peaks result from trajectories in configuration space with only diffusion steps and that trajectories with
reactions make up the part between the peaks, we can conclude that reactions are responsible for the
peaks in the fluctuation ratios. In order to verify this assumption we analyzed the contributions to
the fluctuation ratio coming from the different types of trajectories. We show in Fig. \ref{fig10} that the 
observed minima and maxima are indeed mainly due to the trajectories with a single reaction process.
For this we compare the fluctuation ratio with the quantity $\Pi_F(\phi)/\Pi_R(-\phi)$ where $\Pi(\phi)$ is the probability
distribution for all trajectories having (a) only diffusion steps or (b) exactly one reaction process. Obviously, the peaks in the
latter ratio coincide with the peaks in the fluctuation ratio.

\begin{table}[thb]
\begin{tabular}{|c|c|c|}
\hline 
PD maxima & FR maxima & FR minima \\
\hline
$-1.63$ & $-1.78$ & $-1.5$ \\
$-0.61$ & $-0.72$ & $-0.38$ \\
0.60 & 0.50 & 0.82 \\
2.02 & 1.89 & 2.25 \\
3.64 & 3.44 & 3.84 \\
\hline
\end{tabular}
\caption{Positions of the maxima in the probability distributions (PD) and of the maxima and minima
in the fluctuation ratio (FR) for model 3, with $D=5$, $h_0=0.2$, $\Delta h=1.2$,  
and $\lambda =1$. The system size is $L=8$ and the driving length is $M=6$. \label{table2}}
\end{table}

\begin{figure}[ht]
\epsfxsize=3.00in\ \epsfbox{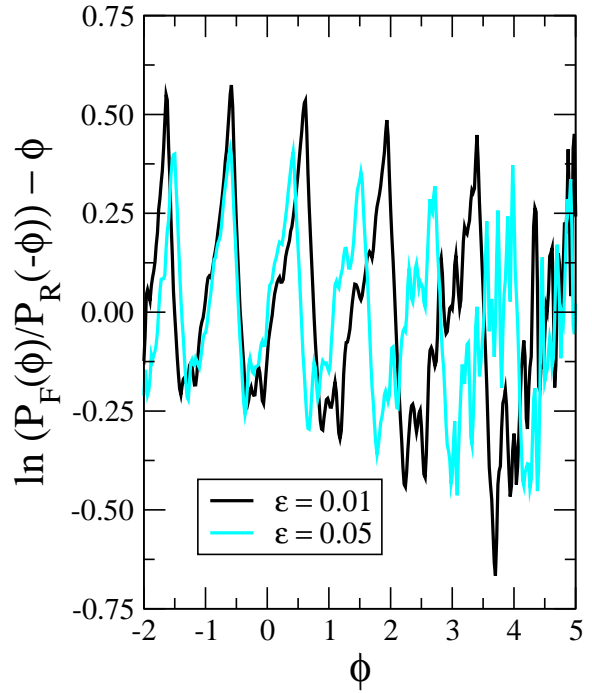}
\caption{(Color online) Fluctuation relations for model 3' and different values of $\varepsilon$. The values
of the parameters are $D=5$, 
$h_0=0.2$, $\Delta h=1.2$, and $\lambda =1$. The system size here is $L=10$ and the driving length is $M=10$. 
These data have
been obtained through Monte Carlo simulations.
}
\label{fig11}
\end{figure}

As a second interesting observation we note that the oscillations in the fluctuation
ratios are not restricted to cases where microscopic reversibility is broken but are much more widespread. As is 
shown in Fig. \ref{fig11} for model 3' (the same holds for model 4') peaks in the fluctuation ratios also show up
in some systems where all reactions are reversible.

\begin{figure}[ht]
\epsfxsize=3.00in\ \epsfbox{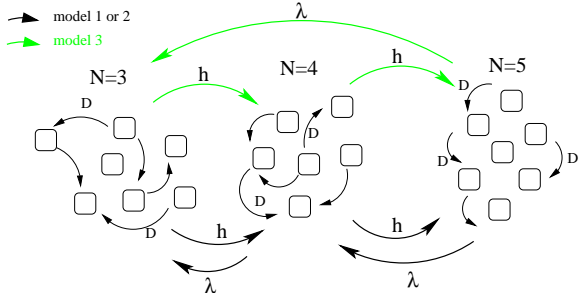}
\caption{(Color online) Schematic plot of the configuration space for models 1, 2, and 3 where configurations are grouped by the number of particles in 
the system $N$. In a diffusion step the system goes from one configuration to another in the same unit without changing the particle number.
Passages between different units are due to reaction processes. For models 1 and 2 one always has that $\Delta N = \pm 1$. This is different
for model 3 (and 4) where different reactions yield different changes in the number of particles in the system.
\label{fig12}}
\end{figure}

In order to understand the origin of these oscillations we need to go back to the different reaction schemes
summarized in Table \ref{table1}. The configuration space of a reaction-diffusion system 
can be thought to be composed of smaller units formed by the configurations with a common number $N$ of particles.
A diffusion step conserves the number of particles, thereby connecting two configurations in the same unit.
A passage from one unit to another always involves a change of particle number and is therefore exclusively 
due to a reaction process. This is sketched in Fig. \ref{fig12}. Keeping this in mind, a fundamental difference emerges between
models 1 and 2 on the one hand and models 3 and 4 on the other hand. In the former systems every reaction changes
the particle number by 1, $\Delta N = \pm 1$. In the latter systems, however,
also larger changes in the particle number happen in the
annihilation process, with
$\Delta N = -2$ for model 3 and $\Delta N = -3$ for model 4. As a consequence, loops in configuration space that connect a unit
with constant $N$ with itself and that involve reactions will display an asymmetry in the number of creation and annihilation processes. 
Thus for model 3 the smallest
loop contains two creation processes and one annihilation. 
This effect is still present, even though
in a weaker form, when we add the backreactions and end up with a microscopically reversible model like model 3' with a variable number of particles
added or subtracted in the different reactions.
It is this difference in the number of particles created in a creation process 
or destroyed in an annihilation event that yields contributions to the probability distributions 
which are not compensated in the fluctuation ratio. 

\section{Discussion}
Characterizing the out-of-equilibrium properties of interacting many-body systems remains one of the most challenging tasks 
in contemporary physics. The recent advent of exact fluctuation and work theorems yielded some excitement in the community
as it indicated a possible way of characterizing large classes of nonequilibrium systems.

In our work we try to characterize diffusion-limited reactions both in their nonequilibrium steady state and in the
transient state when the systems are driven out of stationarity. For systems in their steady state we confirm the expectation
that probability currents allow to distinguish between equilibrium and nonequilibrium steady states. In addition, they also
allow to define a global quantity that quantifies the distance to equilibrium. This way of characterizing nonequilibrium
steady systems remains valid even when microscopic reversibility is broken, as it is the case for many reaction-diffusion
systems.

The situation is more complicated if one wishes to characterize reaction-diffusion systems through fluctuation and work
theorems. If one studies a system for which microscopic reversibility is fulfilled, one can define a work-like quantity,
our quantity $R$, see eq. (\ref{ds}), for which exact detailed fluctuation theorems not only hold in the steady states but are also valid
when the system is driven out of stationarity through time dependent reaction rates. In absence of microscopic reversibility,
however, $R$ can not be used as it is no longer well defined. Instead we propose to use the driving entropy production $\phi$, see eq. (\ref{dphi}),
initially introduced in \cite{Hat01,Esp07}, as this quantity exclusively uses stationary probabilities and therefore remains well defined
even in the absence of microscopic reversibility. Whereas the driving entropy production always fulfills a global fluctuation theorem \cite{Hat01,Esp07}, it 
only fulfills a detailed fluctuation theorem for systems with equilibrium steady states. At first look, this seems to strongly reduce
the usefulness of his quantity for the characterization of systems with nonequilibrium steady states. However, as we showed in this paper, 
the deviations of the fluctuation ratios for $\phi$ from a simple exponential behavior do contain non-trivial information on the
trajectories in configuration space. Indeed, in cases where the change in the number of particles is different for different reactions,
we observe systematic deviations from a simple exponential behavior. These deviations, which take the form of peaks superimposed on an 
exponential, mainly result from trajectories in configuration space where exactly one reaction takes place.

It is not an easy task to quantitatively relate the peak heights and the peak positions to the values of the system parameters.
For this, a much more in-depth study is needed where all the parameters are varied in a systematic way \cite{Dor10}.

Whereas the driving entropy production $\phi$ remains a well-defined quantity even in the absence of microscopic reversibility,
we need to mention that in many cases this could be a quantity that is difficult to measure as the knowledge of
the stationary probability distribution of the system is required. As a consequence, the practical importance of $\phi$
could be restricted, especially for experimental systems where the stationary probability distribution is often not
easily accessible.

How general are the results found in this work? Based on the reaction schemes discussed in this work and given in Table \ref{table1},
we expect the peaks to appear in the fluctuation ratios for $\phi$ for any reaction-diffusion system that allows for a variable number of particles to
be created or destroyed in the different reactions. This also encompasses more complicated systems with two or more particle types.
In addition, signatures of the same type should also be observed for other system classes with a configuration space topology 
that is similar to that of the the reaction-diffusion
systems (i.e., composed by groups of configurations that are only connected in a very specific way) and 
with a similar asymmetry in the 
configuration space trajectories. An extension of our work along these lines is planed for the future.

\begin{acknowledgments}
We thank Chris Jarzynski and Fr\'{e}d\'{e}ric van Wijland for interesting and stimulating discussions.
This work was supported by the US National
Science Foundation through DMR-0904999.
\end{acknowledgments}

\section*{Appendix}
In order to compute the probability distribution of $\phi$, see (\ref{dphi}), when changing some rate 
$r$ from its initial value $r_0$ to the final value $r_M$ in $M$ steps, with $r_i = r_0 + \frac{i}{M} (r_M - r_i )$, $i=0, 
\cdots M$, we first need to know the stationary probability distributions for any value $r_i$. 
This is easily done by determining the null eigenvector of the Liouville matrix. We then need 
to generate all possible sequences
of configurations ({\it paths} in configuration space) ${\bf X} = C_0 \longrightarrow C_1 \longrightarrow \cdots
\longrightarrow C_{M-1} \longrightarrow C_M$, where only one reaction or diffusion takes place at every step.
Starting form every possible initial configuration, we have to built up a tree structure to all the configurations that can be reached
in $M$ steps with non-zero probability. This is done recursively by a standard depth-first search algorithm that ends when we
reach the $M$th step. We now have to attach a probability to every one of these generated paths. For this we are multiplying
the probability to select the initial configuration $C_0$ with the product of the $M$ transition probabilities:
\begin{equation}
P_F\left( {\bf X} \right) = P_s(C_0,r_0) \prod\limits_{i=0}^{M-1} \omega(C_i \longrightarrow C_{i+1},r_{i+1} )~.
\end{equation}
Having now determined every path and its probability, we need in addition the values of $\phi$
along these different paths, which we obtain through the equation
\begin{equation}
\widetilde{\phi}\left( {\bf X} \right) = \sum\limits_{i=0}^{M-1} \left( \ln P_s(C_i,r_{i+1}) - \ln P_s(C_i,r_i) \right)~,
\end{equation}
where $P_s(C_i,r_i)$ is the stationary probability to find the configuration $C_i$ at the value $r_i$ of the rate $r$.
Putting everything together, the probability distribution is finally obtained through the expression
\begin{equation}
P_F(\phi) = \sum\limits_{\bf X} P_F\left( {\bf X} \right) \, \delta \left( \widetilde{\phi} \left( {\bf X} \right) - \phi \right)~.
\end{equation}
In addition to the just described forward process we also study the reversed process where we start in configuration
$C_M$ with the rate $r_M$ before changing the reaction rate in $M$ steps to its final value $r_0$. The
probability distribution for this process is then
\begin{equation}
P_R(\phi) = \sum\limits_{\bf \widetilde{X}} P_R\left( {\bf \widetilde{X}} \right) \, \delta \left( \widetilde{\phi} \left( {\bf \widetilde{X}} \right) - \phi \right)
\end{equation}
with
\begin{equation}
P_R\left( {\bf \widetilde{X}} \right) = P_s(C_M,r_M) \prod\limits_{i=0}^{M-1} \omega(C_{M-i} \longrightarrow C_{M-1-i},
r_{M-1-i})~.
\end{equation}

In Section III we discuss not only the quantity $\phi$ but also the quantity $R$ defined by Eq. (\ref{ds}). 
For this second
quantity the procedure is exactly the same, only the calculation of the values of $\phi$ for the different paths has to be
replaced by the values of $R$.

\begin{figure}[thb] \label{fig13}
\centerline{\epsfxsize=3.00in\ \epsfbox{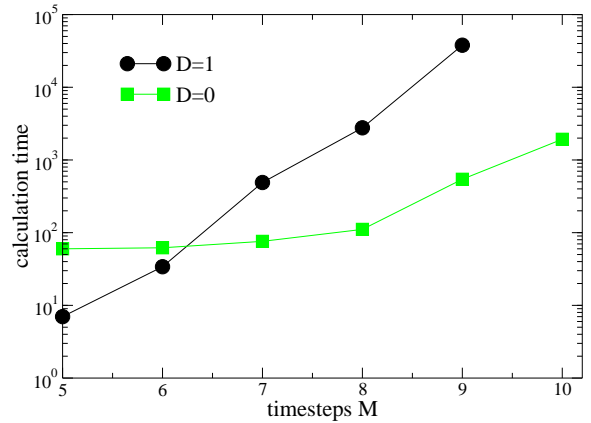}}
\caption{(Color online) Exponential growth of the calculation time in function of the number of steps
for model 1 with $L=6$ sites where the creation rate $h$ was changed between $h_0 =0.2$ and $h_M = 1.4$. 
For this calculation we set $\lambda=1.0$ and considered both vanishing ($D=0$) and non-vanishing ($D=1$) diffusion rates.}
\end{figure}

This numerical exact approach is limited to small system sizes $L$ and few steps $M$, as the number of
paths grows exponentially with both $L$ and $M$, see Fig. \ref{fig13}. 
For example, for $L= 6$ the number of paths increases from 404 for
$M=2$ to 8.6 $10^8$ for $M=9$. 

\end{document}